\documentclass[12pt]{article}
\usepackage{graphicx}
\usepackage{amsmath}
\usepackage{amsfonts}
\usepackage{amssymb}
\newtheorem{theorem}{Theorem}

\begin{document}

\title{First-order transitions for $n$-vector models in two and more dimensions;
rigorous proof.}
\author{Aernout C. D. van Enter\\Institute for Theoretical Physics\\Rijksuniversiteit Groningen \\P.O. Box 800\\9747 AG Groningen \\the Netherlands\\aenter@phys.rug.nl
\and Senya. B. Shlosman\\CPT, CNRS Luminy, Case 907 \\F13288 Marseille Cedex 9 France\\shlosman@cptsu5.univ-mrs.fr}
\maketitle

{\bf Abstract: We prove that various SO(n)-invariant $n$-vector models with  
interactions which have
%is sufficiently steep and narrow 
a deep and narrow enough minimum
have a first-order transition  
in the temperature. The result holds in dimension two or more, and
is independent on the nature of the low-temperature phase.}

Recently Bl\"{o}te, Guo and Hilhorst \cite{BGH}, extending earlier work by
Domany, Schick and Swendsen \cite{DSS} on 2-dimensional classical XY-models,
performed a numerical study of 2-dimensional $n$-vector models with non-linear
interactions. For sufficiently strong values of the non-linearity, they found
the presence of a first-order transition in temperature. In \cite{DSS} a
heuristic explanation of this first-order behavior, based on a similarity with
the high-$q$ Potts model, was suggested, explaining the numerical results. A
further confirmation of this transition was found by Caracciolo and Pellisetto
\cite{CP}, who considered the $n\rightarrow\infty$ (spherical limit) of the
model, and found the same first-order transition.

On the other hand, various studies, mostly based on Renormalization-Group
analyses or Kosterlitz-Thouless type arguments based on the picture of
binding-unbinding of vortices, have contested this first-order behaviour (e.g.
\cite{him,kno,jon}).

Here we settle the issue by presenting a rigorous proof of the existence of
this first-order transition.

It may seem somewhat surprising that two-dimensional $n$-vector models, whose
magnetization by the Mermin-Wagner theorem \cite{MW} is always zero, can have
such a phase transition. The reason is that the transition we are talking
about here is manifested by the long-range order in higher-order correlation
functions. Such transitions were discovered by one of us some time ago, see
\cite{Sh1}. But the results of \cite{Sh1} were related to the fact that  there
the symmetry group was the (disconnected) group $O\left(  2\right)  ,$ and at
the transition point the discrete symmetry $\mathbb{Z}_{2}$ is broken, while
the connected part -- $SO\left(  2\right)  $ -- of the symmetry persists. The
nature of the transition we study here, however, is not connected to any type
of symmetry breaking, and as such is much closer to the high-$q$ $q$-state
Potts model, or the model studied in \cite{DS}, where a first order transition
between a low-energy and a high-entropy phase occurs. Thus we confirm the
original intuition of \cite{DSS}.

For XY-spins in two dimensions there can be a low-temperature phase with slow
polynomial decay of correlations, while the majority belief in the field,
despite the work of Patrascioiu and Seiler \cite{PS}, is that for $n>2$, the
$n$-vector models at low finite temperatures have exponentially decaying
correlations, just as at high temperatures. Our result unfortunately does not
say anything about this question.

Our proof is directly inspired by the existing proofs for
low-energy--high-entropy phase transitions, and is indeed an adaptation of
those. We employ the method of Reflection Positivity \cite{RP} (RP). For
simplicity we write the proof for 2-dimensional XY-spins, the extension to the
general case is immediate.

We remark that also the generalization to higher dimensions is immediate. Thus
the low-temperature phase can be either magnetized, Kosterlitz-Thouless-like
(not magnetized with slow correlation decay), or possibly non-magnetized with
exponentially decaying correlations. As such our result contradicts strong
``universality'' claims, stating that universality classes of interactions
exist, all elements of which have the same kind of phase transition between a
high-temperature phase and a low-temperature phase, and which are only
determined by the dimension of the system, the symmetry of the interaction and
whether the interaction is short-range or long-range.

We will consider the Hamiltonian given by
\begin{equation}
H=-J\sum_{i,j\in\mathbb{Z}^{2}}\left(  \frac{1+\cos\left(  \phi_{i}-\phi
_{j}\right)  }{2}\right)  ^{p}.\label{21}%
\end{equation}
To formulate our result we have to introduce for every n.n. bond $b=\left(
i,j\right)  $ the following bond observables:
\begin{equation}
P_{b}^{<}\left(  \phi_{i},\phi_{j}\right)  =\left\{
\begin{array}
[c]{ll}%
1 & \text{ if }\left|  \phi_{i}-\phi_{j}\right|  <\varepsilon/2,\\
0 & \text{ if }\left|  \phi_{i}-\phi_{j}\right|  >\varepsilon/2,
\end{array}
\right.  \label{22}%
\end{equation}
which project on the ordered bond configurations, and $P_{b}^{>}\left(
\phi_{i},\phi_{j}\right)  =1-P_{b}^{<}\left(  \phi_{i},\phi_{j}\right)  .$ Our
main result is contained in the following

\begin{theorem}
Suppose the parameter $p$ is large enough. Then there exists 
%the 
%{\bf 
a
%}
transition
temperature $\beta_{c}=\beta_{c}\left(  J,p\right)  ,$ such that there are two
different Gibbs states, $\left\langle \cdot\right\rangle ^{<}$ and
$\left\langle \cdot\right\rangle ^{>},$ at $\beta=\beta_{c},$ corresponding to
the Hamiltonian (\ref{21}).  For some specific choice of $\varepsilon
=\varepsilon\left(  p\right)  $ in (\ref{22}), we have for the ``ordered''
state $\left\langle \cdot\right\rangle ^{<}$ that
\[
\left\langle P_{b}^{<}\right\rangle ^{<}>\kappa\left(  p\right)  ,
\]
while in the ``disordered'' phase 
%$\left\langle \cdot\right\rangle ^{<}$%
$\left\langle \cdot\right\rangle ^{>}$%
%{\bf I changed < to > in the superscript}
\[
\left\langle P_{b}^{>}\right\rangle ^{>}>\kappa\left(  p\right)  ,
\]
for each bond $b,$ with $\kappa\left(  p\right)  \rightarrow1$ as
$p\rightarrow\infty.$
\end{theorem}

Before analysing the model (\ref{21}), we  present an even simpler toy model,
which already displays the mechanism, and which is even closer to the Potts
model. The single spin space is the circle, $S^{1},$ the free measure is the
Lebesgue measure, normalized such that $S^{1}$ has measure one, so
$S^{1}=\left[  -\frac{1}{2},\frac{1}{2}\right]  $. ( One can take here any
sphere $\mathbb{S}^{n}$ instead.) The toy Hamiltonian is
\begin{equation}
H=-J\sum_{i,j\in\mathbb{Z}^{2}}U\left(  \phi_{i},\phi_{j}\right)
\end{equation}
The nearest neighbour interaction $U\left(  \phi_{1},\phi_{2}\right)
=U\left(  \left|  \phi_{1}-\phi_{2}\right|  \right)  $ is given by
\[
U\left(  \phi\right)  =\left\{
\begin{array}
[c]{ll}%
-1 & \text{ if }\left|  \phi\right|  \leq\frac{\varepsilon}{2},\\
0 & \text{otherwise.}%
\end{array}
\right.
\]
Here $\varepsilon$ plays a similar role to $\frac{1}{q}$ in the $q$-state
Potts model. The Hamiltonian is RP under reflections in coordinate planes. For
the case $\mathbb{Z}^{2}$ one has also RP under reflections in lines at 45
degrees, passing through the lattice sites. That case is the easiest.

One has to show that:

\begin{itemize}
\item $\left\langle P_{b}^{>}\right\rangle _{\beta}$ is small for large
$\beta;$ (the ordered, typical low-temperature phase bonds)

\item $\left\langle P_{b}^{<}\right\rangle _{\beta}$ is small for small
$\beta;$ (the disordered, typical high-temperature phase bonds)

\item $\left\langle P_{b^{\prime}}^{<}P_{b^{\prime\prime}}^{>}\right\rangle
_{\beta}$ is small for all $\beta,$
\end{itemize}

\noindent provided $\varepsilon$ is small enough. Here $\left\langle
\cdot\right\rangle _{\beta}$ is the state with periodic b.c. in the box
$\Lambda$ of size $L,$ and $b^{\prime},b^{\prime\prime}$ are two orthogonal
bonds sharing the same site. The estimates have to be uniform in $L,$ for $L$
large. The first two are straightforward application of RP and the chess-board
estimate. So let us get the last one. By the chess-board estimate,
\begin{equation}
\left\langle P_{b^{\prime}}^{<}P_{b^{\prime\prime}}^{>}\right\rangle _{\beta
}\leq\left\langle P^{\#}\right\rangle _{\beta}^{1/\left|  \Lambda\right|
},\label{01}%
\end{equation}
where the observable $P^{\#}$ is the indicator
\[
P^{\#}=\prod_{b\in E_{01}}P_{b}^{<}\prod_{b\in E_{23}}P_{b}^{>}.
\]
Here $E_{01},E_{23}$ is the partition of all the bonds in $\Lambda$ into two
halves; $E_{01}$ consists of all bonds $\left(  x,x+e_{1}\right)  $ and
$\left(  x,x+e_{2}\right)  ,$ for which $x^{1}+x^{2}=0$ or
$1\mathrm{\operatorname{mod}\,}4,$ while $E_{23}$ is the other half; $e_{1}$
and $e_{2}$ are the two coordinate vectors.

To proceed with the estimate (\ref{01}) we need the estimate on the partition
function. We have
\begin{equation}
Z_{\Lambda}\left(  \beta,\varepsilon\right)  \geq\left(  \frac{\varepsilon}%
{2}\right)  ^{\left|  \Lambda\right|  }e^{2\beta\left|  \Lambda\right|
}+\left(  1-4\varepsilon\right)  ^{\left|  \Lambda\right|  /2}.\label{03}%
\end{equation}
(The first summand is obtained by integrating over all configurations $\phi,$
such that $\left|  \phi_{x}\right|  \leq\frac{\varepsilon}{4}$ for all
$x\in\Lambda.$ For the second one we take all configurations $\phi$ which are
arbitrary on the even sublattice and which satisfy $\left|  \phi_{x}-\phi
_{y}\right|  >\varepsilon/2$ for every pair of n.n.; for every $y$ on the odd
sublattice that leaves the spins to be free in a set of measure $\geq
1-4\varepsilon.$) Solving
\[
\left(  \frac{\varepsilon}{2}\right)  ^{\left|  \Lambda\right|  }e^{2\beta
_{0}\left|  \Lambda\right|  }=\left(  1-4\varepsilon\right)  ^{\left|
\Lambda\right|  /2}%
\]
for $\beta_{0}$ we find
\begin{equation}
e^{4\beta_{0}}=\left(  1-4\varepsilon\right)  \left(  \frac{\varepsilon}%
{2}\right)  ^{-2},\label{02}%
\end{equation}
so for $\beta\geq\beta_{0}$ the first term in (\ref{03}) dominates, while for
$\beta\leq\beta_{0}$ the second term dominates. Similarly, the partition
function $Z_{\Lambda}^{\#}\left(  \beta,\varepsilon\right)  ,$ taken over all
configurations $\phi$ with $P^{\#}\left(  \phi\right)  =1$ satisfies
\begin{equation}
%\[
Z_{\Lambda}^{\#}\left(  \beta,\varepsilon\right)  \leq e^{\beta\left|
\Lambda\right|  }\varepsilon^{\frac{3}{4}\left|  \Lambda\right|  +O\left(
\sqrt{\left|  \Lambda\right|  }\right)  }.
%\].
\label{88}%
\end{equation}
If $\beta\geq\beta_{0},$ we write, using (\ref{02}):
\[
\left\langle P_{b^{\prime}}^{<}P_{b^{\prime\prime}}^{>}\right\rangle _{\beta
}\leq\frac{e^{\beta}\varepsilon^{\frac{3}{4}}}{\frac{\varepsilon}{2}e^{2\beta
}}=2\frac{1}{\varepsilon^{\frac{1}{4}}e^{\beta}}\leq2\frac{1}{\left[
\varepsilon\left(  1-4\varepsilon\right)  \left(  \frac{\varepsilon}%
{2}\right)  ^{-2}\right]  ^{\frac{1}{4}}}\leq C\varepsilon^{1/4}.
\]
If $\beta\leq\beta_{0},$ we similarly have
\[
\left\langle P_{b^{\prime}}^{<}P_{b^{\prime\prime}}^{>}\right\rangle _{\beta
}\leq\frac{e^{\beta}\varepsilon^{\frac{3}{4}}}{\left(  1-4\varepsilon\right)
^{1/2}}\leq\frac{\left[  \left(  1-4\varepsilon\right)  \left(  \frac
{\varepsilon}{2}\right)  ^{-2}\varepsilon^{3}\right]  ^{\frac{1}{4}}}{\left(
1-4\varepsilon\right)  ^{1/2}}\leq C^{\prime}\varepsilon^{1/4}.
\]
So we are done.

%\textbf{This last paragraph is either too long or too short, and I think that
%it has to be longer. 
% You may be right, but if we try Phys Rev Lett, 
%we are subjected to a strict upper bound for the length} 

For the non-linear models, we employ the fact that for small difference angles
$\cos(\phi_{i}-\phi_{j})$ is approximately $1-O((\phi_{i}-\phi_{j})^{2})$ and
furthermore that $\lim_{p\rightarrow\infty}(1-\frac{1}{p})^{p}=e^{-1}$.
%\textbf{(Where this is used?)} 
%{\bf 
This suggests to choose $\epsilon(p) = \frac{1}{\sqrt{p}}$. Because the
separation between ordered and disordered bonds is somewhat arbitrary, to 
obtain an inequality similar to (5), we make 
a slightly different choice.
%}
We consider a bond $(i,j)$ disordered if
$|\phi_{i}-\phi_{j}|\geq\frac{C}{\sqrt{p}}$ for some large $C$. So first we
choose a sufficiently large constant $C$. For the estimate of the ordered
partition function we only integrate over the much smaller intervals of
``strongly ordered'' configurations: $|\phi_{i}|\leq\frac{C^{-1}}{\sqrt{p}}$
to obtain a lower bound: 
%{\bf 
\begin{equation} 
Z_{\Lambda}\left(  \beta,p\right)  \geq\left(  
%\frac{\varepsilon}{2} 
\frac{1}{2 C \sqrt p} \right)  ^{\left|  
\Lambda\right|  }e^{[2\beta(1 - O(\frac{1}{C}))]\left|  
\Lambda\right| }+\left(  1-\frac{4C}{\sqrt p}\right)  ^{\left|  
\Lambda\right|  /2}.\label{09}% 
\end{equation}
%}

This 
%guarantees 
%{\bf 
makes use of the fact
%}
that the ``strongly ordered'' bonds
all have energy almost equal to $-J$, whereas the disordered partition
function is bounded by that of the toy model, 
%{\bf 
but
%} 
with $\varepsilon$ replaced by
$\frac{C}{\sqrt{p}}$. 

%{\bf 
For the estimate which shows that ordered and disordered bonds tend not
to neighbor each other, we obtain:

\begin{equation} 
Z_{\Lambda}^{\#}\left(  \beta,p\right)  \leq e^{\beta\left| \Lambda\right|  }
({\frac{C}{\sqrt p}})^{(\frac{3}{4}+ O(e^{-C}))\left|  
\Lambda\right|  +O\left( \sqrt{\left|  \Lambda\right|  }\right)  }. 
\end{equation}
%}

%{\bf Remove:(As asymptotically $\frac{ln\frac{1}{\sqrt{p}}}%
%{ln\frac{C^{2}}{\sqrt{p}}}$)} 

%\textbf{(where this expression comes from?) }
%{\bf (This is because I was thinking of 
%computing free energies densities, instead of 
%partition functions, ordered $2J$, cq $2J - O(\frac{1}{C})$, disordered
%$-ln \sqrt p$, cq $- ln C \sqrt p$, and bad $\frac{3}{4}$ times that.)} 
%approaches $1$, the argument still works for $p$ sufficiently large.

%{\bf Another way of saying this, which is not good tex, but I 
%am writing this at home at my brother-in-law's is:
%In the inequality (5), in the right hand side, the first term then picks 
%up an extra term ${(C^{-1} \times exp( - O (C^{-1})))}^{\Lambda}$, while 
%the 2nd term  becomes $ (1 - 4 C \sqrt p) exp (O( exp (-C)))^{\Lambda}$ 
%for large p. 
%{\bf The ensuing
%analysis still goes through, with the power $\frac{1}{4}$ replaced 
%by a slightly smaller one. 
%Below is the equation replacing (5)}

%and the equation replacing the other needed inequality is

%{\bf 
The rest of the argument is essentially unchanged. 
We first chose C big enough (such that $\frac{1}{C}$ 
is small wrt to $1$),  and we can still choose $p$ big enough such that $p$ 
is large wrt to $C^2$, which finishes the argument.
%}

{}
\end{document}